\documentclass{aa}  

\usepackage{graphicx}
\usepackage{txfonts}
\usepackage{lipsum}
\usepackage{subcaption}         % necessary for continued figures, example in section 3
\usepackage{lscape}             % to rotate a single page table, example in appendix.
\usepackage{placeins}           % useful with \FloatBarrier, to keep 
\usepackage{xcolor}                 
\usepackage{amsmath, mathtools,derivative}

\newcommand{\mesa}{\texttt{MESA}\xspace}
\newcommand{\mespa}{\texttt{MESPA}\xspace}

\usepackage[print-unity-mantissa = false]{siunitx}
\DeclareSIUnit[]{\kb}{k_B}
\DeclareSIUnit[]{\yr}{yr}
\DeclareSIUnit[]{\Gyr}{\giga\yr}
\DeclareSIUnit{\bary}{m_u}
\DeclareSIUnit{\MJ}{M_J}
\DeclareSIUnit{\ME}{M_E}
\DeclareSIUnit{\RJ}{R_J}
\DeclareSIUnit{\RJeq}{R_J^{eq}}
\DeclareSIUnit{\kbperbary}{\kb \per \bary}
\DeclareSIUnit{\erg}{erg}
\DeclareSIUnit{\bar}{bar}

\newcommand{\Zcore}{Z_\mathrm{core}}
\newcommand{\Zenv}{Z_\mathrm{env}}
\newcommand{\senv}{s_\mathrm{env}^\odot}
\newcommand{\score}{s_\mathrm{core}^\odot}
\newcommand{\mstrat}{m_\mathrm{strat}}
\newcommand{\Zmean}{\overline{Z}}
\newcommand{\mmid}{m_\mathrm{mid}}
\newcommand{\Mp}{M_\mathrm{p}}
\newcommand{\dZ}{\delta Z}
\newcommand{\dP}{\delta P}

\newcommand{\dR}{\delta R}
\newcommand{\dRRel}[1]{\left(\frac{\dR}{R}\right)_{#1}}
\newcommand{\dRRelInline}[1]{(\dR / R_1)_{#1}}
\newcommand{\ds}{\delta s}
\newcommand{\sTh}{s_{\Zmean}}
\newcommand{\dsTh}{\delta \sTh}

\newcommand{\taumature}{\SI{1e9}{\yr}}

\newcommand{\tauKH}{\tau_\mathrm{KH}}
\newcommand{\tauKHDot}{\dot{\tau}_\mathrm{KH}}
\newcommand{\tauKHnod}{\tau_{\mathrm{KH},0}}
\newcommand{\tauadj}{\tau_\mathrm{adj}}
\newcommand{\tauadjnod}{\tau_{\mathrm{adj},0}}
\newcommand{\tauDecouple}{\tau_\mathrm{dec}}

\newcommand{\Eg}{E_\mathrm{g}}
\newcommand{\EgDot}{\dot{E}_\mathrm{g}}
\newcommand{\Eth}{E_\mathrm{th}}
\newcommand{\EthDot}{\dot{E}_\mathrm{th}}

\newcommand{\tauNought}{\SI{1.93e09}{\yr}}
\newcommand{\expTauNought}{9.28}
\newcommand{\expTauNoughtErr}{0.02}
\newcommand{\expMp}{0.71}
\newcommand{\expMpErr}{0.03}
\newcommand{\expZBulk}{-0.46}
\newcommand{\expZBulkErr}{0.05}
\newcommand{\expOpacity}{0.22}
\newcommand{\expOpacityErr}{0.01}
\newcommand{\meanOpacity}{\SI{30.6}{\cm \squared \per \gram}}

\begin{document}

   \title{The influence of composition gradients on giant planet radii}

   \author{
   H.~Knierim\inst{\ref{inst1}} \and
   R.~Helled \inst{\ref{inst1}}
   }

\institute{
Department of Astrophysics, University of Zurich, Winterthurerstrasse 190, CH-8057 Zurich, Switzerland
\email{henrik.knierim@uzh.ch}\label{inst1}
}

\date{Received 27 May 2026 / Accepted 3 July 2026}

\abstract{
The radius of a giant planet is a key physical property. 
It encapsulates the outcome of its mass, composition, and thermal state, serving as a critical link between observations and the theory of planetary interiors.
While traditional interior models assume distinct core--envelope structures, recent gravity measurements of Jupiter and Saturn reveal complex interiors with deep composition gradients.
Here, we combine an analytic framework with numerical evolution simulations to elucidate how internal structures with composition gradients affect the planetary radius.
We demonstrate that the spatial redistribution of heavy elements does not inherently alter the planetary size; rather, it is the resulting difference in total entropy over time that drives observable radius variations.
We show that this mechanism naturally establishes two distinct evolutionary regimes. 
During the first few $\SI{1e9}{\yr}$ of evolution, composition gradients trap thermal energy deep within the interior, significantly affecting the planetary evolution and internal structure. 
However, as the planet cools and contracts, this ``thermal memory'' fades and the total entropy no longer depends on the primordial conditions. 
Consequently, the planetary radius decouples from the internal distribution of heavy elements, becoming solely a function of the total mass and bulk composition.
Our results present the physical origin of the mass--radius--composition relation and suggest that the use of simplified interior models for giant exoplanets is legitimate only after the planet has surpassed this evolutionary threshold.
Our analysis establishes a thermodynamic constraint on the planetary structure: while composition gradients can temporarily modulate cooling, they cannot permanently sustain radius inflation, as the addition of heavy elements, regardless of their distribution, inevitably leads to planetary contraction.
}


   \maketitle
   \nolinenumbers

%%%%%%%%%%%%%%%%%%%%%%%%%%%%%%%%%%%%%%%%%%%%%%%%%%%%%%%%%%%%%%
\section{Introduction}
The radius of a giant planet is our most accessible probe of the physics governing its deep interior.
As a macroscopic observable, it integrates the complex interplay of mass, composition, and cooling history into a measurable quantity.
\par
Consequently, modeling giant planet radii has a long tradition. 
The first mass--radius relations were derived using polytropes \citep[e.g.,][]{Chandrasekhar1939} and then refined through numerical integrations of cold homogeneous spheres \citep{Zapolsky_1969}.
Subsequent work moved beyond the zero-temperature approximation \citep[see][for a review]{Stevenson1982}, and later models relaxed the isentropic assumption to construct non-adiabatic interiors \citep[e.g.,][]{Guillot1994}. These developments established the canonical framework of giant planet interior theory, where a gas giant consists of a heavy-element core surrounded by an inner metallic and outer molecular hydrogen and helium (H-He) envelope.
\par
With the discovery of exoplanets with accurately measured masses and sizes, it became essential to develop simple yet robust interior models. Historically, the vast majority of giant exoplanets were analyzed by adopting the layered and isentropic structures mentioned above, an approach that readily enables statistical studies across entire populations \citep[e.g.,][]{Swift2012, Weiss2014, Batygin2013, Mueller2024}. A direct consequence of interpreting this wealth of mass--radius data through these simplified models is the prominent mass--metallicity relation \citep[e.g.,][]{Guillot2008, Miller2011, Thorngren2016, Mueller2025, Chachan2025}, which serves as a critical constraint on planet formation theories \citep[e.g.,][]{Shibata_2020, Schneider2021, Knierim_2022b}. 
\par
At the same time, gravity measurements by the Juno and Cassini missions revealed that Jupiter and Saturn possess fuzzy cores \citep{Wahl2017, Iess2019}. The Solar System's gas giants are therefore neither neatly layered nor fully isentropic. Instead, their interiors demand careful treatment of equations of state (EoS), thermal profiles, and other physical phenomena such as helium phase separation \citep[see][and references therein]{Howard2025}. This realization motivated a new generation of evolutionary frameworks that incorporate dilute cores, convective mixing, and more realistic atmospheres \citep[e.g.,][]{Vazan2018, Mueller2020, Sur2024, Knierim2024, Tejada_Arevalo2025, Helled_2025, Knierim2026}.
\par
Yet, despite these advances, exoplanet studies still largely rely on simplified, layered models. While recent studies included composition gradients \citep[e.g.,][]{van_Dijk2025, Peerani2026}, convective mixing is often neglected, leading to an overestimate of the planetary radius \citep{Knierim2025a}. 
By explicitly including this mixing, we demonstrated that primordial structures can in fact be constrained from observations. Specifically, a single mass--radius measurement can yield different bulk metallicities depending on the planet's initial composition gradient and thermal state \citep[i.e., potential entropy; see][]{Knierim2024}.
This raises the question of whether the widely used mass--metallicity relation, derived assuming distinct core--envelope structures, remains robust or requires re-evaluation.
\par
Here, we investigate the role of composition gradients in shaping gas giant radii. We combine an analytic perturbation framework with numerical simulations to isolate the underlying drivers of planetary inflation and contraction. By comparing models of identical mass and bulk composition, but with different composition gradients and primordial thermal states, we explicitly separate the thermal, pressure, and compositional contributions over time. 
Our paper is organized as follows. In Sect.~\ref{sec:analytic_model} we derive the analytic radius perturbation and demonstrate how the structural memory of internal composition gradients is lost over time. To test this analytic prediction, Sect.~\ref{sec:numerical_methods} details our numerical methods. We then highlight the observational implications in Sect.~\ref{sec:observational_implications}, and present our discussion and conclusions in Sect.~\ref{sec:discussion}.
%%%%%%%%%%%%%%%%%%%%%%%%%%%%%%%%%%%%%%%%%%%%%%%%%%%%%%%%%%%%%%

%%%%%%%%%%%%%%%%%%%%%%%%%%%%%%%%%%%%%%%%%%%%%%%%%%%%%%%%%%%%%%
\section{Analytic model}\label{sec:analytic_model}
In the cold, degenerate limit, the radius of a gas giant is determined by its mass and bulk composition, largely independent of its internal stratification \citep{Zapolsky_1969, Stevenson2025}. However, while the planet is still contracting and cooling, its interior structure depends intimately on its local specific entropy profile, which is shaped by both primordial formation conditions and the presence of composition gradients.
To demonstrate how different internal heavy-element distributions drive variations in planetary radii, we formulate an analytic model based on perturbation theory.
Formally, the internal composition is described by the set of independent mass fractions $\{X_i\}$ for $N$ chemical species (where $i = 1, \dots, N-1$).
To streamline the analysis, we aggregate all heavy elements into the metallicity $Z$ and assume a constant hydrogen-to-helium ratio ($X/Y$).
This reduces the composition to a one-parameter function of $Z$. The mathematical structure of the derivation remains analogous for the general case involving the full set of species.

Consider two planetary models of identical mass $M$ but distinct internal metallicity profiles, denoted $Z_1$ and $Z_2$. We construct both profiles to yield the same bulk metallicity $\Zmean$.
Consequently, the integral of their difference must vanish,
\begin{align}\label{eq:bulk_composition_conserved}
    \int_{0}^{M}\delta Z(m)\odif{m} = 0,
\end{align}
where $\dZ(m) = Z_2(m) - Z_1(m)$ represents the composition perturbation at mass coordinate $m$.
To connect these composition differences to a radius perturbation, we begin by recalling that the volume of model $i$ is given by the integral of its specific volume $v_i$,
\begin{align}\label{eq:mass_conservation}
   \frac{4}{3} \pi R_i^3 = \int_{0}^{M} v_i(m) \odif{m}.
\end{align}
Since we are ultimately interested in the radius discrepancy $\dR = R_2 - R_1$, we expand $R_2^3 \approx R_1^3 + 3R_1^2\dR$, assuming that $\dR$ is small.
Substituting Eq.~\eqref{eq:mass_conservation} into this expansion allows us to express the radius offset directly in terms of the specific volume contrast between the two models
\begin{align}\label{eq:dR_full}
    \dR = \frac{1}{4 \pi R_1^2}\int_{0}^{M} \left[ v_2(m) - v_1(m) \right] \odif{m}.
\end{align}

To first order, the difference in specific volume is given by the total differential
\begin{align} \label{eq:first_order_approx}
    v_2 - v_1 \approx \pdv{v}{s}_{P,Z} \ds + \pdv{v}{P}_{s,Z} \dP + \pdv{v}{Z}_{s,P} \dZ,
\end{align}
where $\ds = s_2 - s_1$ and $\dP = P_2 - P_1$ are the specific entropy and pressure differences, respectively.
To meaningfully compare the internal states of two planets with different composition gradients, we must account for the fact that specific entropy inherently changes with composition. Thus, we decompose the specific entropy perturbation $\ds$ into two distinct parts: a composition-adjusted ``thermal'' specific entropy difference $\dsTh$, which represents the offset if both models were homogeneously mixed to the bulk metallicity $\Zmean$, and a compositional shift $\lambda \dZ$, where $\lambda$ accounts for the specific entropy difference caused solely by the local addition of heavy elements. 
By substituting $\ds = \dsTh + \lambda \dZ$ back into Eq.~\eqref{eq:first_order_approx}, we can group all direct composition dependencies
\begin{align} \label{eq:differential_grouped}
    v_2 - v_1 \approx \pdv{v}{s}_{P,Z} \dsTh + \pdv{v}{P}_{s,Z} \dP + \zeta \dZ,
\end{align}
where $\zeta \coloneqq \pdv{v}/{s}_{P,Z} \lambda + \pdv{v}/{Z}_{s,P}$.
Inserting this expanded differential into Eq.~\eqref{eq:dR_full} yields three distinct fractional radius contributions: a thermal term $\dRRelInline{s}$, a pressure term $\dRRelInline{P}$, and a unified composition term $\dRRelInline{Z}$. 
\par
Inside gas giants, the effective specific volume response to composition $\zeta$ is nearly uniform with mass (see Appendix~\ref{sec:validity_analytic} for details). This uniformity allows us to extract it from the integrand, giving 
\begin{align}\label{eq:partial_propto_rho}
    \dRRel{Z} \approx \frac{\zeta}{4 \pi R_1^3} \int_{0}^{M} \dZ \odif{m}.
\end{align}
The integral on the right-hand side is exactly the difference in total metal mass between the two models, which is---by construction---zero. Consequently, the direct radius perturbation arising from the composition gradient vanishes.
\par
\begin{figure}
    \centering
    \includegraphics[width=1.0\linewidth]{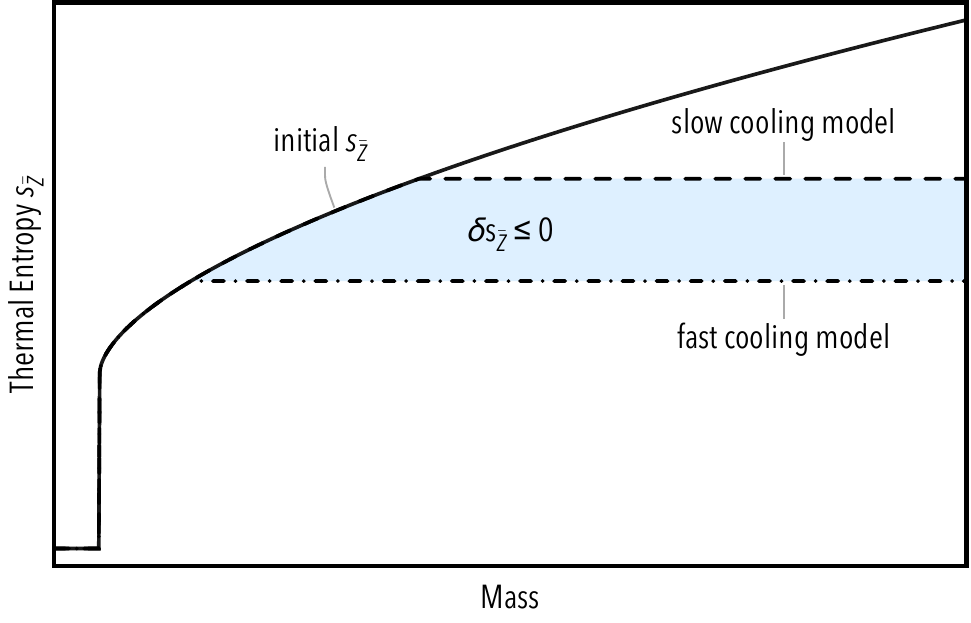}
    \caption{A sketch of the thermal entropy, $\sTh$, as a function of mass. The dashed and dash-dotted lines represent two models cooling from an identical initial $\sTh$ profile (solid curve), but with different composition profiles. The shaded area highlights that the resulting difference, $\dsTh$, maintains a constant sign.}
    \label{fig:dS_thermal_single_signed}
\end{figure}

To evaluate the specific entropy term, we consider two models starting with identical thermal specific entropy profiles. Since gas giants lose heat (ignoring anomalous heating), any thermal specific entropy difference between them maintains a constant sign (for a sketch, see Fig.~\ref{fig:dS_thermal_single_signed}). Assuming without loss of generality that the second model cools faster, $\dsTh(m) \leq 0$ throughout the interior. Evaluating these objects when their globally integrated thermal entropies match ($\int_0^M \dsTh \odif{m} = 0$) imposes a strict mathematical constraint. Because a function of constant sign integrates to zero only if $\dsTh(m) = 0$ everywhere, the thermal radius perturbation vanishes.
\par
If planets instead form with disparate thermal specific entropy profiles, $\dsTh$ may initially alternate in sign. In this scenario, positive and negative $\dsTh$ regions could integrate to zero (without the function vanishing locally). This can theoretically lead to a residual thermal radius perturbation. In practice, however, most of these primordial thermal specific entropy differences reside in the outer envelope. Because composition gradients typically do not extend to this outer region, it becomes fully convective and loses its thermal memory on a Kelvin--Helmholtz timescale.
As a result, $\dsTh$ is expected to effectively evolve to a state of a constant sign, recovering the mathematical constraint that leads to a vanishing radius perturbation. 
\par
Assuming hydrostatic equilibrium, the pressure perturbation scales as
\begin{align} \label{eq:pressure_perturbation}
    \dP(m) = -\int_{m}^{M} \frac{G m}{4 \pi r^4} \left(-4 \frac{\delta r}{r}\right) \odif{m} = - 4\frac{\dR}{R_1} P(m),
\end{align}
where $r$ is the radial coordinate and we applied the homology approximation $\delta r / r = \dR / R_1$ (see Appendix \ref{sec:validity_analytic} for details). Substituting Eq.~\eqref{eq:pressure_perturbation} into Eq.~\eqref{eq:dR_full} yields
\begin{align} \label{eq:P-term}
    \dRRel{P} = \frac{1}{4 \pi R_1^3} \left(4 \frac{\delta R}{R_1} \right)\int_{0}^{M} \frac{v}{\Gamma_1} \odif{m},
\end{align}
where we incorporate the thermodynamic identity $\pdv{v}/{P}_{s,Z} = - v / (\Gamma_1 P)$, with $\Gamma_1$ representing the first adiabatic index. For two planets with the same bulk metallicities and total entropies, the pressure term constitutes the entire fractional radius difference ($\dRRelInline{P} = \dR/R_1$). Thus, we can factor out the radius perturbation to obtain
\begin{align}
    0 = \frac{\dR}{R_1}\left(1 - \frac{4}{3C}\right),
\end{align}
where $C$ is a structural constant. This relation presents two mathematical solutions. If $\Gamma_1$ is uniform throughout the interior, $C = \Gamma_1 = 4/3$. This value corresponds to the well-known limit for dynamical instability, representing the exact threshold where the radius of a fluid sphere decouples from its mass \citep{Chandrasekhar1939}. Because stable gas giants require $\Gamma_1 > 4/3$, this first branch is physically inaccessible. Thus, the only viable solution is a vanishing macroscopic radius perturbation ($\dR = 0$).
\par
Synthesizing these derivations reveals that, for a given mass and bulk metallicity, the size of a gas giant is determined by its total entropy, $S = \int_{0}^{M} s(m) \odif{m}$. This thermodynamic constraint holds independently of the internal distribution of heavy elements $Z(m)$ and specific entropy $s(m)$. Local variations in these profiles do not perturb the planetary radius, provided their integrals are conserved and the entropy offset maintains a constant sign.
This thermodynamic re-framing establishes two distinct evolutionary regimes. During the ``entropy-divergent'' phase, composition gradients strongly influence cooling by trapping heat deep within the interior and altering the opacity throughout the planet. Consequently, the total entropies of stratified and homogeneous models diverge significantly at a given age, driving substantial radius discrepancies. However, as the planets cool and degeneracy increasingly governs the internal structure, their total entropies naturally converge. In this ``entropy-convergent'' phase, the direct composition perturbation vanishes to first order. Consequently, the planetary radius decouples from the internal heavy-element distribution, becoming a function solely of total mass and bulk metallicity.
\section{Numerical experiments}\label{sec:numerical_methods}
\begin{figure}
    \centering
    \includegraphics[width=1.0\linewidth]{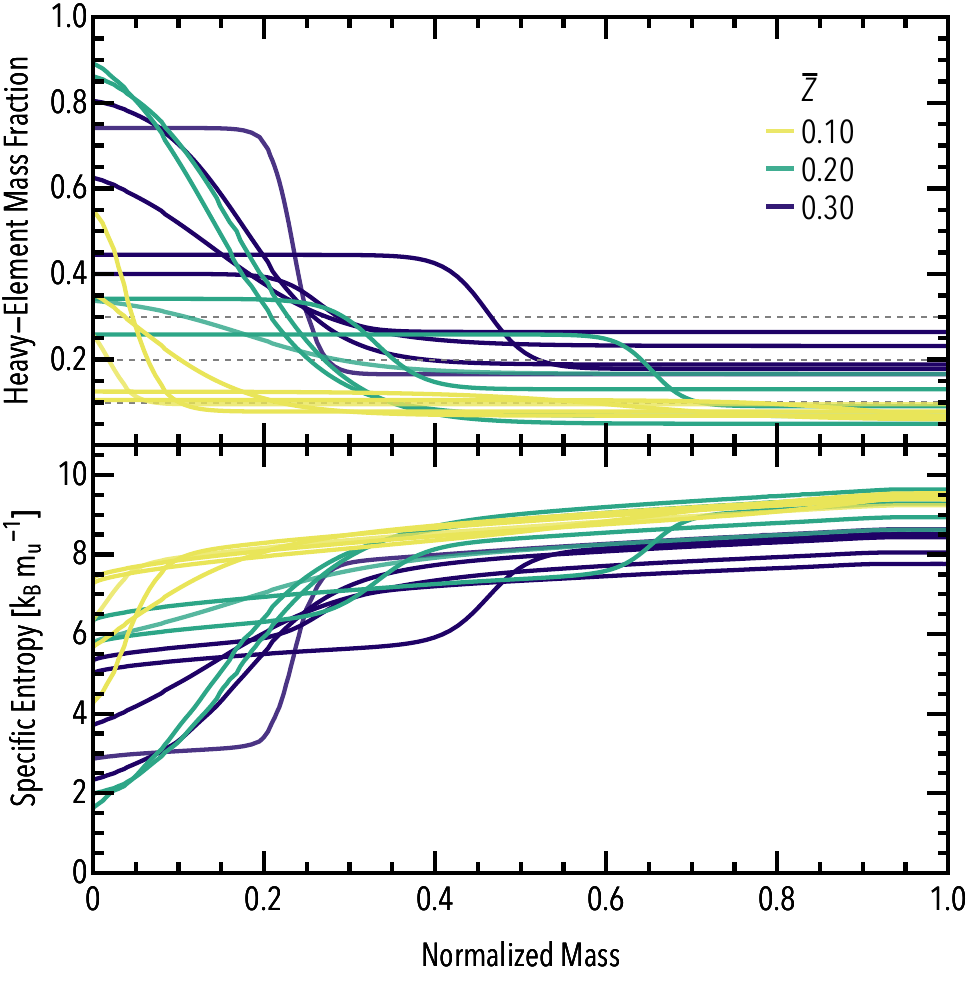}
    \caption{Composition (top) and specific entropy (bottom) profiles of randomly generated initial conditions.}
    \label{fig:initial_profile_example}
\end{figure}

To scrutinize the validity of our analytic approximation, we performed a suite of simulations covering planetary masses of $\Mp = \SI{0.5}{\MJ}$, \SI{1.0}{\MJ}, and \SI{2.0}{\MJ}, each with bulk metallicities of $\Zmean = 0.1$, 0.2, and 0.3.
The planetary evolution was modeled using the \mesa code \citep{Paxton_2011, Paxton_2013, Paxton_2015, Paxton_2018, Paxton_2019,Jermyn_2023}, modified for planets \citep[\mespa, see][and references therein]{Helled_2025}.
We employed the H-He EoS of \citet{Chabrier_2021} and represented the metallicity $Z$ as a half-half mixture of rock and water \citep[for details, see][]{Mueller2020}.
The composition profiles were drawn from random distributions but constrained such that, within each simulation run, they reproduced a prescribed bulk metallicity (see Appendix~\ref{sec:initial_model_setup} for details).
Example curves are shown in Fig.~\ref{fig:initial_profile_example}.
For the atmosphere, we employed the semi-gray atmosphere model from \citet{Guillot2010} using Jupiter's equilibrium temperature. Finally, the models were evolved for \SI{1e11}{\yr}---a timescale exceeding the age of the Universe, chosen to ensure the planets settle into their final, cold-degenerate configuration.

\begin{figure*}
    \centering
    \includegraphics[width=1\linewidth]{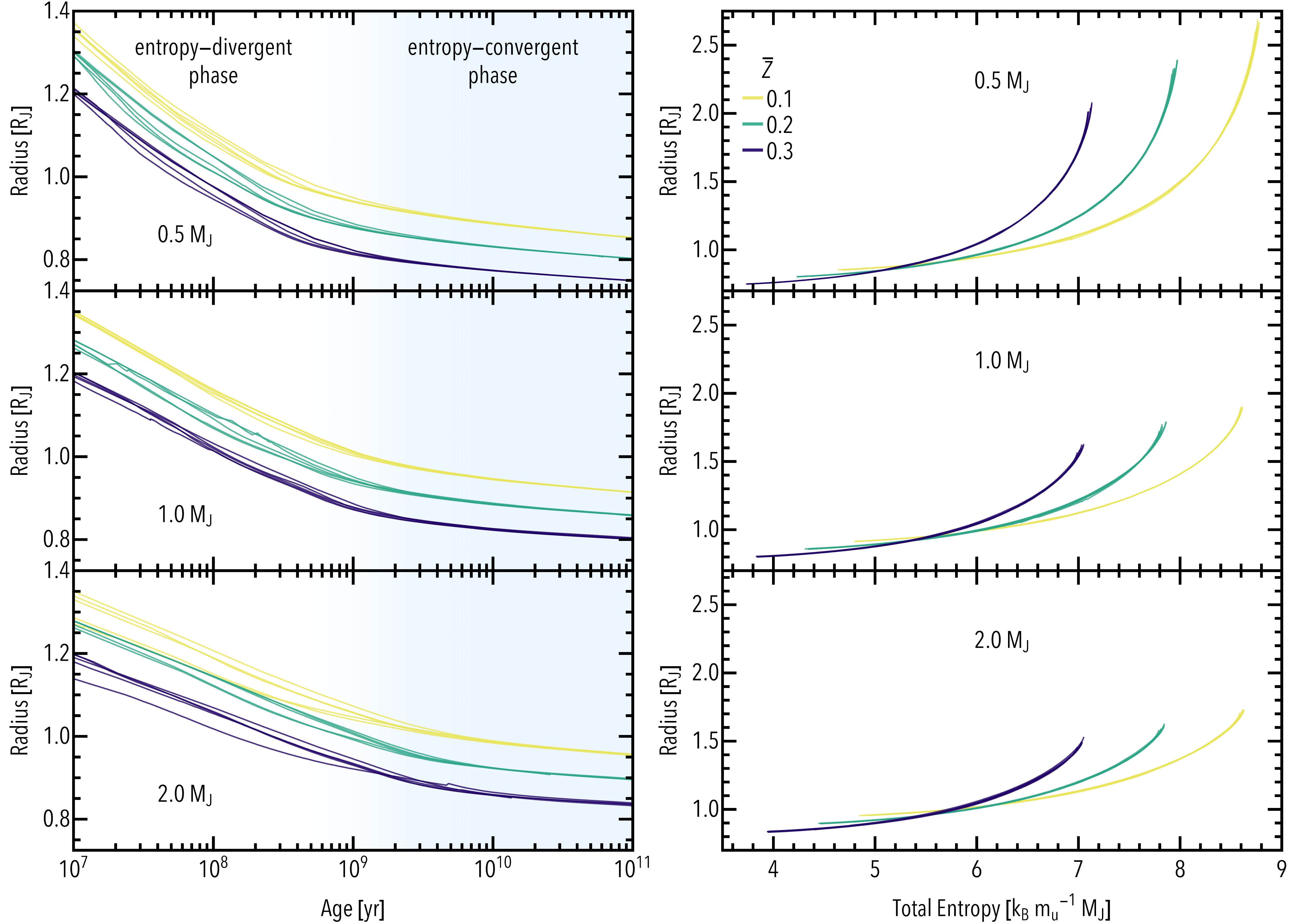}
    \caption{Planetary radius as a function of time (left) and total entropy (right) for models with different composition gradients.}
    \label{fig:radius_evolution}
\end{figure*}

The results, displayed in Fig.~\ref{fig:radius_evolution}, confirm the predicted behavior.
At ages $\lesssim$ a few $\taumature$, planets with identical bulk properties but different composition gradients can exhibit noticeable radius discrepancies.
However, as the planets cool and evolve toward an entropy-convergent state, their radii converge. Conversely, when parameterized by total entropy, the divergent evolutionary tracks perfectly overlap, directly validating the derivation in Sect.~\ref{sec:analytic_model}.
\subsection{The decoupling timescale}
The time it takes for a planet to reach the entropy-convergent phase depends on the thermal evolution, and more specifically, the material functions.
If the planetary interior is hot and the composition gradient is shallow, thus mixes rapidly, convection can efficiently cool off any formation differences, leading to the well-known radius convergence on the order of a Kelvin--Helmholtz timescale $\tau_\mathrm{KH}\sim \SI{1e7}{\yr}$ \citep[e.g.,][]{Knierim2025a}.
However, colder interiors with steep composition gradients resist mixing \citep{Knierim2024}, preserving radiative regions that are stable against convection.
There, heat transport is diffusive and operates on the much longer thermal adjustment timescale \citep[e.g.,][]{Kippenhahn2012}
\begin{equation} \label{eq:tau_adj}
    \tauadj \approx \frac{c_v}{\overline\sigma^*}\mstrat^2 \sim \SI{4e10}{\yr},
\end{equation}
where $c_v \sim \SI{7e7}{\erg\per\gram\per\K}$ is the mean specific heat capacity, $\overline\sigma^* \sim \SI{2e48}{\erg\gram\per\s\per\K}$ is the mean effective conductivity, and $\mstrat \sim \SI{0.1}{\MJ}$ is the mass of the stably stratified region.
Although the region that is stable against convection dissipates its internal heat over $\tauadj$, entropy convergence occurs earlier.
\par
Specifically, the total entropy converges once the thermal adjustment timescale of the core equals the Kelvin--Helmholtz timescale of the envelope, $\tauadj(\mstrat) = \tauKH$.
As long as $\tauKH < \tauadj$, the convective envelope cools faster than the radiative layer. Consequently, it induces a negative potential entropy gradient, which initiates convection \citep[see ``effective entropy'' in][]{Knierim2024}. Therefore, once $\tauKH = \tauadj$, both cooling rates become comparable and no further convection can be triggered.
\par
This process is self-regulating. If the primordial core forms very hot, it is entirely eroded by convective mixing, leading to convergence on the short Kelvin--Helmholtz timescale. Conversely, a very cold primordial core would have a negligible contribution to the total entropy of the planet. Therefore, there is a sweet-spot, where cores are sufficiently hot to affect the radius evolution, but are also sufficiently cold to persist. Numerically, we find that these ``warm'' cores influence the radius evolution only during the first $\sim\SI{1e9}{\yr}$. We define this age, where the total entropies converge and the radius thus decouples from the interior distribution of heavy elements, as the ``decoupling age''.
\par
Analytically, we derive the associated decoupling timescale by assuming that the thermal adjustment timescale for a specific mass coordinate remains roughly constant in time, $\tauadj(\mstrat) \approx \tauadjnod(\mstrat)$. In contrast, the Kelvin--Helmholtz timescale scales approximately linearly with time, $\tauKH \approx \tauKHDot t + \tauKHnod$, where the dot denotes the time derivative. This derivative expands to $\tauKHDot = L^{-1} \EgDot - L^{-2} \Eg \dot{L}$, with $L$ representing luminosity and $\Eg$ the gravitational energy. Because the first term is much smaller than the second, it can be neglected. Given that the luminosity of a gas giant originates primarily from its thermal energy \citep[$L \approx -\EthDot$, e.g.,][]{Stevenson2025}, and assuming a power-law relation $L\propto \Eth^\gamma$, we obtain
\begin{equation}
    \tauKHDot \approx \frac{\Eg}{\Eth} \gamma.
\end{equation}
Equating the thermal adjustment timescale of the core to the Kelvin--Helmholtz timescale yields the decoupling timescale
\begin{equation}
    \tauDecouple = \frac{\tauadjnod(\mstrat)-\tauKHnod}{\gamma\Eg/\Eth} \approx \frac{\tauadjnod(\mstrat)}{\gamma} \frac{\Eth}{\Eg},
\end{equation}
where we neglect $\tauKHnod$ since $\tauKHnod \ll \tauadjnod$. Substituting the stable core mass from Eq.~\eqref{eq:tau_adj} and assuming a typical initial cooling derivative of $\tauKHDot \sim 40$, we infer $\tauDecouple \sim \SI{1e9}{\yr}$, which is in excellent agreement with the numerical results.
\par
Although this analytic formulation elegantly captures the underlying key physics, its practical application depends on several assumptions and requires a priori knowledge of the stable core mass $\mstrat$ \citep[for an analytic model to compute $\mstrat$, see][]{Knierim2024}. We therefore also extracted an empirical timescale directly from our simulations (for details, see Appendix \ref{sec:power-law_details}). By fitting a power law to the epoch where the radius discrepancy between the stratified and homogeneous models drops to \qty{1}{\percent} for planetary masses between $0.3$ and \SI{3}{\MJ}, we derive the decoupling relation
\begin{multline}\label{eq:decoupling_timescale}
    \tauDecouple \approx \tauNought \left(\frac{\Zmean}{0.1}\right)^{\expZBulk} \left(\frac{\overline\kappa}{\meanOpacity}\right)^{\expOpacity} \left(\frac{M_\mathrm{p}}{\SI{1}{\MJ}}\right)^{\expMp},
\end{multline}
where $\overline\kappa$ is the mean reference opacity evaluated at an age of \SI{1e7}{\yr}.
\par
To physically interpret these empirical exponents, we can compare them to the thermal adjustment timescale. Invoking homology, Eq.~\eqref{eq:tau_adj} simplifies to $\tauadj \propto \kappa \Mp^{2/3}$. Our empirically derived mass exponent approximately reproduces this theoretical expectation ($\expMp \approx 2/3$). However, for the opacity, the scaling deviates significantly from this baseline. Instead of a direct linear dependence ($\tauDecouple \propto \kappa$), we find a strongly sub-linear relationship ($\expOpacity \ll 1$). 
\par
This sub-linear scaling stems from the complex relationship between heavy-element distribution and the local opacity. Because the opacity typically increases with metallicity, a planet with composition gradients possesses a more transparent envelope and a more opaque deep interior in comparison to its homogeneous equivalent. Consequently, the cooling of the atmosphere and the decrease of the envelope's entropy occurs more rapidly for the compositionally stratified model during the early evolution. The homogeneous object eventually catches up once the envelope of the planet with composition gradients has already cooled because the metal-rich deep interior still retains its primordial heat.
\par
Scaling the baseline opacity profile by a constant factor $f_\kappa$ delays global cooling. Simultaneously, however, this multiplication amplifies the absolute opacity contrast between the two configurations.
This introduces a competing effect: although increased opacities naturally prolong cooling and yield larger radii at a given age, they also force the models to converge at a higher total entropy, and therefore at a larger absolute radius. This offset effectively dampens the temporal dependence on $\kappa$, producing the observed sub-linear behavior.
\par

Finally, as long as the planetary deep interior remains stably stratified, the bulk composition leads to a relatively small correction. Because $\tauDecouple$ only depends weakly on the bulk metallicity ($\propto \Zmean^{\expZBulk}$), the decoupling epoch shifts by merely a factor of two even when considering a relatively large range of metallicities ($\Zmean = 0.1$ to $0.5$, see the bottom panel in Fig.~\ref{fig:power_law_fit} and its relevant discussion). Also note that this empirical scaling breaks down in the metal-poor limit (e.g., $\Zmean \lesssim 0.05$). At such low metallicities, the available heavy elements are insufficient to maintain a stabilizing composition gradient against convection. The interior consequently mixes and rapidly homogenizes, causing the radius discrepancy to vanish on the much shorter global Kelvin--Helmholtz timescale $\tau_\mathrm{KH}$.
%%%%%%%%%%%%%%%%%%%%%%%%%%%%%%%%%%%%%%%%%%%%%%%%%%%%%%%%%%%%%%

\section{Connection to exoplanets} \label{sec:observational_implications}
To place our theoretical decoupling limit within the current observational landscape, we compared our derived timescale with the observed exoplanet population. 
We extracted a sample of giant planets with masses between \SI{0.5}{\MJ} and \SI{2.0}{\MJ}, along with their host star ages, from the PlanetS catalog \citep{Parc_2024}.

\begin{figure*}
    \centering
    \includegraphics[width=1\linewidth]{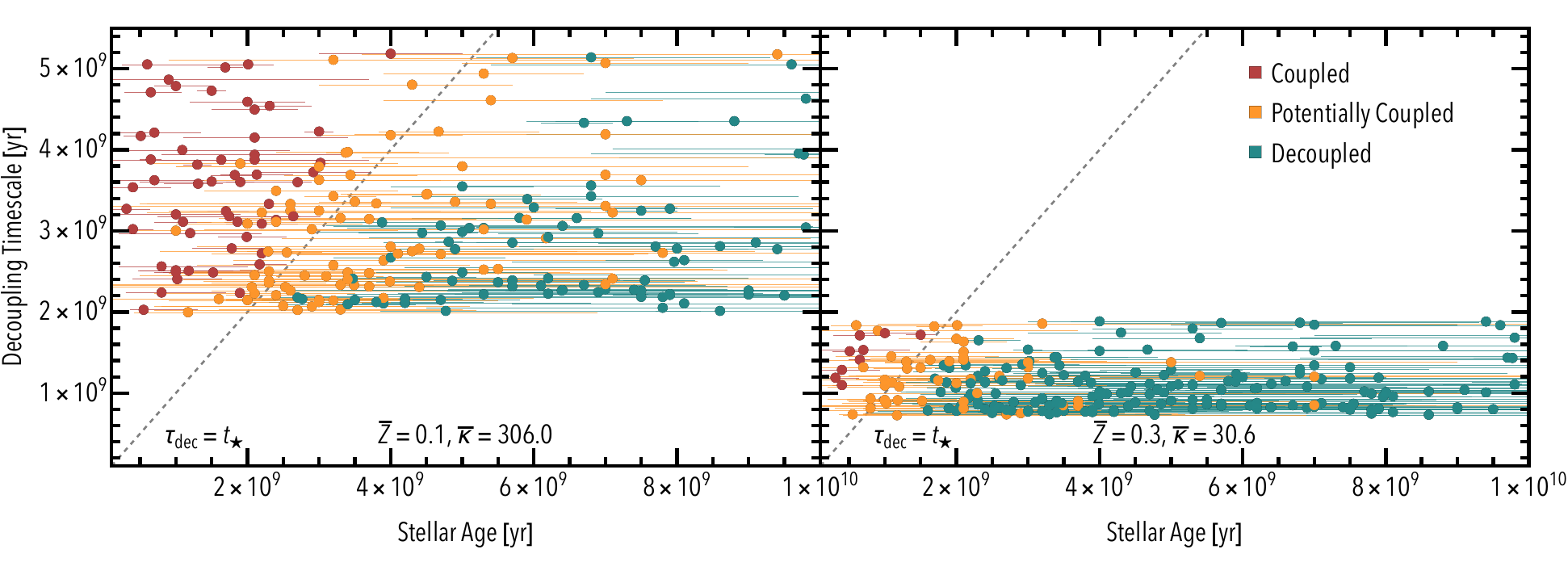}
    \caption{Decoupling timescale $\tauDecouple$ versus stellar age $t_\star$ for giant exoplanets (\SI{0.5}{\MJ} to \SI{2.0}{\MJ}) from the PlanetS catalog \citep{Parc_2024}. The panels show how parameter choices affect the timescale of the transition: rapid (right) and delayed (left) transition. Planets are categorized into ``coupled'', ``potentially coupled'', and ``decoupled'', depending on whether the host star's age estimate (including uncertainties) is strictly younger than, straddles, or is strictly older than $\tauDecouple$.}\label{fig:decoupling_population}
\end{figure*}

For each planet, we computed $\tauDecouple$ using Eq.~\eqref{eq:decoupling_timescale}. To demonstrate the sensitivity of the threshold to the assumed parameters across the population, we evaluated two limiting scenarios. The rapid transition case assumes a metal-rich, nominal-opacity interior ($\Zmean = 0.3$, $\overline\kappa = \meanOpacity$), while the delayed transition case adopts a metal-poor, highly opaque interior ($\Zmean = 0.1$, $\overline\kappa = 10 \times \meanOpacity$). Figure~\ref{fig:decoupling_population} contrasts the resulting distributions. 

The vast majority of the observed giant exoplanets have estimated ages between \SI{1}{\Gyr} and \SI{10}{\Gyr}, coinciding squarely with the decoupling timescale. Rather than residing safely in the entropy-convergent limit, many of these planets may be transitioning out of their entropy-divergent phase. Figure~\ref{fig:decoupling_population} shows this variance and the resulting shift in the transition, demonstrating that the categorization of planets into coupled or potentially coupled regimes changes substantially depending on the parameters used to evaluate Eq.~\eqref{eq:decoupling_timescale}. 
\par
Therefore, specifically for gas giants closer to the \SI{1}{\Gyr} limit, detailed modeling of composition gradients is still required to properly characterize the planetary interior. Toward an age of about \SI{10}{\Gyr}, we can expect that the transition has occurred. In the entropy-convergent regime, the inferred mass--metallicity relations should no longer depend on the assumed composition profile \citep{Peerani2026}. 
\par
Furthermore, while the planetary radius eventually loses its memory of the heavy-element distribution, the broader observational signature remains. A composition gradient alters the internal density profile and permanently sequesters metals in the deep interior. Consequently, vital observables such as the atmospheric metallicity $Z_\mathrm{atm}$ and the Love number $k_2$ remain inextricably linked to the internal stratification, even in the degenerate limit \citep{Knierim2025a,Knierim2026}.
Given that both Jupiter and Saturn likely host fuzzy cores, composition gradients may be a common outcome of giant planet formation \citep{Lozovsky2017,Helled2017}. If population studies continue to rely exclusively on simplified interior structures because they adequately reproduce the late-age radii, they will fail to capture the underlying heavy-element distributions necessary to inform and constrain gas giant formation models.
Thus, while composition gradients cannot permanently alter the planetary radius, they remain a significant, non-negotiable component of giant planet characterization across all evolutionary epochs.
\subsection{Connection to directly imaged planets}
We note that the application of this convergence limit requires knowledge of both the planetary mass and radius. For directly imaged gas giants, these quantities are typically not measured directly, but are instead inferred from the observed luminosity.
Even if the planetary mass could be constrained by independent measurements, the relationship between luminosity and bulk composition remains highly degenerate (see Appendix~\ref{sec:luminosity_evolution} and references therein).
Furthermore, given that directly imaged planets are predominantly young, they generally reside in the entropy-divergent regime ($t \lesssim \tauDecouple$).
Consequently, inferring the bulk metallicity of a directly imaged planet from its mass, age, and luminosity would depend on the assumed internal structure. As long as the planetary radius is not measured independently, this structural degeneracy prevents a unique determination of bulk metallicity for these systems.
\subsection{Connection to highly irradiated planets}
In our fiducial numerical experiments, we adopted the equilibrium temperature $T_\mathrm{eq}$ of Jupiter. However, the observed exoplanet population spans a vast range of incident stellar fluxes, from ultra-hot Jupiters ($T_\mathrm{eq} \gtrsim \SI{2000}{\K}$) to widely separated, directly imaged planets with negligible irradiation. 
\par
As detailed in Appendix~\ref{sec:equilibrium_temperature}, the decoupling timescale exhibits a non-monotonic dependence on the equilibrium temperature due to competing atmospheric and interior cooling effects.
In general, the decoupling timescale tends to increase by a factor of $\sim 2$ for highly irradiated planets, emphasizing that Eq.~\eqref{eq:decoupling_timescale} is an order-of-magnitude estimate rather than a precise prediction. While the exact age of the decoupling depends on the specific condition, the expectation that this transition occurs is robust.
\section{Discussion and conclusions} \label{sec:discussion}
The robustness of our central claim---that two planets of identical mass and bulk composition evolve toward the same radius---depends on the behavior of the EoS. Specifically, this convergence requires $\zeta$ to remain roughly uniform throughout the planetary interior. If H-He and heavy-element mixtures form exotic compounds that significantly alter this derivative, the direct composition contribution $\dRRelInline{Z}$ would no longer vanish. However, if the mixture behaves reasonably well, the decoupling of the planetary radius from the internal heavy-element distribution remains physically sound.
\par
The exact EoS and specific chemical makeup of the deep interior remain uncertain, and the results clearly depend on the adopted EoS. Also, our choice of a half-half mixture of rock and water serves as an idealization for what is likely a complex blend of various species. If this deep-interior mixture were more volatile-rich, composition gradients would erode more easily. Conversely, a more refractory-rich mixture would render these gradients more resilient against convective mixing \citep[e.g.,][]{Knierim2024}.
Because the decoupling timescale is governed by the erosion of the composition gradients and the overall efficiency of energy transport, adopting a different chemical composition would alter the estimate for the decoupling timescale. However, the potential effect of an alternative composition is somewhat absorbed in the existing framework: the effect of compositional changes to the opacity are already captured by the parameterization in Eq.~\eqref{eq:decoupling_timescale}. Changes in the chemical compositions of the composition gradients can be considered by a modification of the planetary potential entropy as discussed in \citet{Knierim2024}.
Overall, while these material uncertainties are expected to shift the timing of the radius decoupling, they do not alter the underlying physical behavior.
\par
Another physical process that can influence the late-stage evolution of gas giants is helium phase separation. At certain pressures and temperatures, helium becomes immiscible in metallic hydrogen, forming droplets that precipitate toward deeper layers. This ``helium rain'' releases energy, altering both the compositional makeup and the thermal structure of the planet \citep[e.g.,][]{Stevenson_1977a, Fortney_2003, Howard2024, Sur_2025}.
Direct measurements of helium depletion in Jupiter's atmosphere, coupled with Saturn's luminosity excess, strongly suggest that this process actively occurs in nature \citep[see][and references therein]{Helled_2024}.
To investigate the impact of helium rain on the decoupling timescale, in Appendix~\ref{sec:helium_rain}, we compare models that include this process against the nominal simulations from Fig.~\ref{fig:radius_evolution}.
\par
We find that helium phase separation increases the planetary radius within \SI{10}{\Gyr} for planets with masses $\Mp \lesssim \SI{1}{\MJ}$. 
However, for the cases we considered, the onset of this radius inflation occurs well after our predicted decoupling timescale (Eq.~\eqref{eq:decoupling_timescale}).
In other words, by the time helium rain begins to significantly alter the radius, the radii of planets with the same mass and bulk composition have already converged.
Remarkably, because the envelopes of the models at a given mass and bulk composition rain out in a similar fashion, this late-stage radius increase is uniform across models with the same mass and bulk composition.
Although the process of helium rain increases the absolute planetary radius, the individual models remain converged. 
Excluding helium rain in models of lower-mass gas giants systematically underestimates their absolute radii, which can lead to an underestimate of the bulk metallicity of low-mass giant exoplanets \citep[see also][]{Sur_2026}.
Although helium phase separation undoubtedly alters the absolute cooling track, the fundamental theoretical framework presented in Sect.~\ref{sec:analytic_model} remains intact, and the principle of late-time radius convergence still holds.
\par
Furthermore, the numerical models used to map the analytical framework onto the planetary evolution are over-simplified. They rely on the parametrization of complex processes such as convection, utilize opacities that carry inherent theoretical uncertainties, and depend on the incident stellar flux (Appendix~\ref{sec:equilibrium_temperature}).
In addition, we adopted comparable primordial (thermal) specific entropy profiles across all numerical experiments. As presented in Sect.~\ref{sec:analytic_model}, the sign-changes induced in $\dsTh$ from varying these initial conditions largely decay on the relatively short Kelvin--Helmholtz timescale, leaving the ultimate correlation between planetary radius and total entropy unaffected.
Although all these factors undoubtedly introduce errors in predicting the absolute radius at a specific age, and hence the decoupling age Eq.~\eqref{eq:decoupling_timescale}, they are expected to merely shift the timing---rather than prevent the onset---of radius convergence. 

In summary, our findings offer the following primary insights into the nature of gas giant radii:
\begin{itemize}
    \item The spatial redistribution of heavy elements within a gas giant does not inherently induce a radius change. Rather, composition gradients alter the thermal evolution; it is the resulting divergence in total entropy over time that drives observable radius variations. 
    
    \item While composition gradients significantly govern the thermal evolution and cooling history of giant planets, this influence is transient. Once a gas giant cools and enters the entropy-convergent phase ($\gtrsim \tauDecouple$, typically at ages of a few gigayears), the radius is determined by its total mass and bulk metallicity, erasing any structural memory of the internal heavy-element distribution.
    
    \item It is thermodynamically impossible to permanently inflate a gas giant by adding heavy elements. For entropy-divergent gas giants, enhanced metallicity can temporarily delay cooling via increased opacities, briefly stalling contraction. However, for an old and cold (entropy-convergent) gas giant, any addition of metals strictly increases the mean density and decreases the radius \citep[see also][]{Zapolsky_1969}. A metal-rich gas giant always contracts to a smaller size than a less enriched counterpart with an identical mass.
    
    \item Mass--metallicity relations for entropy-divergent gas giants ($\lesssim \tauDecouple$) must explicitly account for composition gradients. Conversely, inferred bulk metallicities for entropy-convergent gas giants are insensitive to the assumed internal structure, justifying the use of simplified, layered models for the oldest planetary populations.

     \item Because the ages of most observed giant planets coincide closely with the decoupling timescale, many are actively transitioning between these thermodynamic regimes. Consequently, accurately interpreting their bulk properties requires the explicit treatment of composition gradients. 
\end{itemize}
By synthesizing data from next-generation exoplanetary missions and ground-based surveys with advanced theoretical frameworks, we are poised to better characterize giant planets and uncover the fundamental physics governing their interiors and evolutionary paths.

\begin{acknowledgements}
H. K. thanks Erik Petigura for inspiring him to pursue this work. We thank Konstantin Batygin and Morris Podolak for valuable discussions.
This work has been carried out within the framework of the National Centre of Competence in Research PlanetS supported by the Swiss National Science Foundation under grant \texttt{51NF40\_205606}. The authors acknowledge the financial support of the SNSF.
\end{acknowledgements}

\bibliographystyle{bibtex/aa}
\bibliography{references}
\begin{appendix}
\section{Specific volume response} \label{sec:validity_analytic}
The analytical framework presented in Sect.~\ref{sec:analytic_model} relies on the assumption that the specific volume response to composition variations, $\zeta$, remains uniform across the planet. To test the validity of this assumption, we compute the direct compositional radius perturbation $\dRRelInline{Z}$ for the numerical models from Sect.~\ref{sec:numerical_methods}. If $\zeta$ were strictly constant, this term would completely vanish.

\begin{figure}[ht!]
  \centering
  \includegraphics[width=\columnwidth]{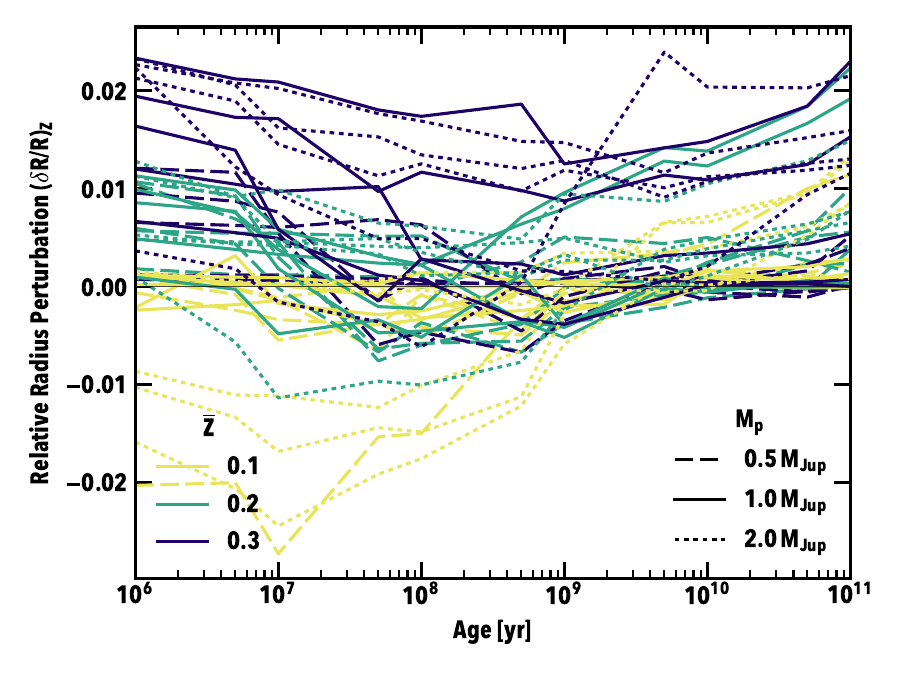}
  \caption{Direct compositional radius perturbation, $\dRRelInline{Z}$, for the evolutionary tracks presented in Fig.~\ref{fig:radius_evolution}.}
  \label{fig:dR_div_R_Z_comparison}
\end{figure}

Figure~\ref{fig:dR_div_R_Z_comparison} demonstrates that this isolated compositional radius discrepancy is on the order of \qty{1}{\percent}.
This confirms that the variation of $\zeta$ is a second-order effect. The full analytic framework also relies on the homology approximation $\delta r/r = \dR/R_1$ to evaluate the pressure term $\dRRelInline{P}$. Because real planets are not perfect, self-similar polytropes, this approximation introduces an additional, concurrent residual to the planetary radius. Crucially, this pressure-driven residual is of comparable magnitude but often acts in the opposite direction.
Together, these relatively small, often counteracting residuals justify our approximation that the planetary radius is decoupled from the internal composition gradient at the same total entropy.
%%%
%%%
%%%
\section{Initial model setup}\label{sec:initial_model_setup}
We constructed the specific entropy and metallicity profiles following the approach of \citet{Knierim2026}. We prescribed the specific entropy before relaxing the composition gradient, i.e., at proto-solar composition $(X, Y, Z)_\odot$, by a power-law profile,
\begin{align}
s^\odot(m) = \score + \left(\frac{m}{M}\right)^{\alpha_s} \cdot (\senv - \score),
\end{align}
where $\score$ and $\senv$ denote the central and surface entropies, respectively, and $\alpha_s$ sets the slope between them.
For the \SI{1}{\MJ} models, we adopted $\score = \SI{8}{\kbperbary}$, $\senv = \SI{10}{\kbperbary}$, and $\alpha_s = 0.6$, yielding specific entropy curves consistent with the detailed calculations of \citet{Cumming2018}. To ensure convergent initial conditions across the mass range, we scaled the envelope entropies for the \SI{0.5}{\MJ} and \SI{2}{\MJ} models to $\senv=\SI{9.5}{\kbperbary}$ and $\SI{11}{\kbperbary}$, respectively.
\par
We drew the composition profiles randomly from a generalized logistic function,
\begin{align}
Z(m) = \Zcore - \frac{\Zcore-\Zenv}{1+\exp\left[{\alpha_Z} (m - \mmid)\right]},
\end{align}
where $\Zcore$ is sampled uniformly between 0 and 1, $\Zenv$ uniformly between 0 and $\min(\Zmean,\Zcore)$, and $\alpha_Z$ log-uniformly between 1 and 100. We then adjusted the midpoint $\mmid$ such that the resulting profile has exactly the desired bulk metallicity $\Zmean$. Profiles that do not satisfy this condition were discarded. The hydrogen-to-helium ratio was set to the protosolar value.
\par
While this functional form inherently limits the diversity of possible composition gradients---for instance, the profile is always symmetric around $\mmid$---and cannot capture all complex outcomes of planet formation, it serves as a practical compromise. Crucially, the logistic curve allows us to smoothly sample the parameter space between classical, sharp core-envelope boundaries and highly extended, dilute cores.
\par
In all the numerical experiments, we first constructed an isentropic model of the specified mass and protosolar composition. We then relaxed the specific entropy profile using \texttt{relax\_initial\_entropy} and subsequently imposed a composition gradient with \texttt{relax\_initial\_composition}.
\section{Power-law fit for the decoupling timescale} \label{sec:power-law_details}
\begin{figure}[ht!]
    \centering
    \includegraphics[width=\columnwidth]{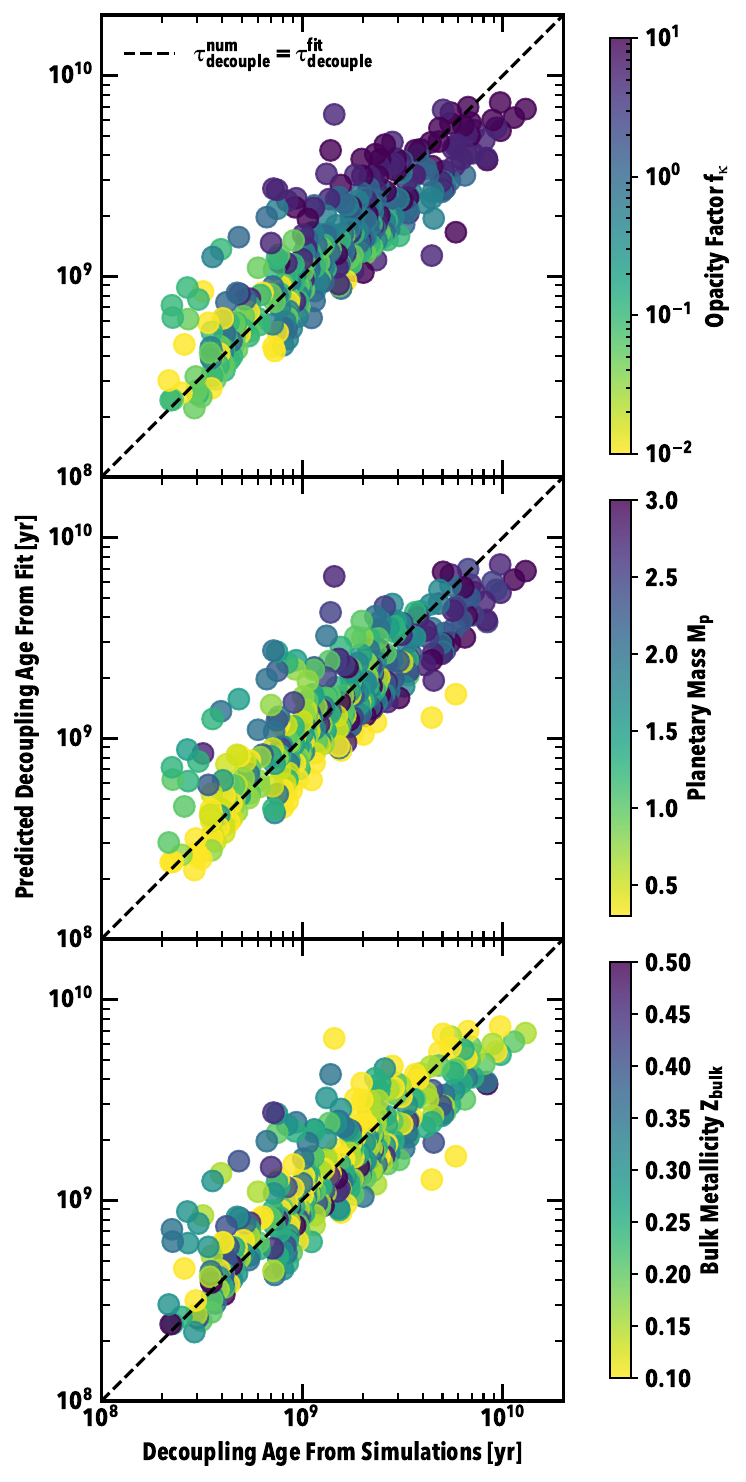}
    \caption{Comparison between the numerically derived convergence ages and the predictions from Eq.~\eqref{eq:decoupling_timescale}.}
    \label{fig:power_law_fit}
\end{figure}

To agnostically determine the scaling of the decoupling timescale, we generated a comprehensive grid of evolution models. 
Specifically, we varied the planetary mass from $\Mp = \SI{0.5}{\MJ}$ to \SI{3.0}{\MJ} in increments of \SI{0.25}{\MJ}, with an additional model at \SI{0.3}{\MJ} to capture the Saturn-mass regime. The bulk composition was varied from $\Zmean = 0.1$ to 0.5 in steps of 0.05.
In addition, we scaled the opacity by a factor $f_\kappa \in \{0.01, 0.05, 0.1, 0.5, 1, 5, 10\}$.
For each run, we extracted the age at which the macroscopic radius of a stratified planet converges to within \qty{1}{\percent} of its homogeneous counterpart.
Prior to fitting, we normalized the grid parameters to reference values: a mass of $\Mp = \SI{1}{\MJ}$, a bulk metallicity of $\Zmean = 0.1$, and the mean opacity (by mass) of $\overline\kappa = \meanOpacity$ (evaluated at \SI{1e7}{\yr} for the unmodified $f_\kappa = 1.0$ case).
We then performed a linear regression in log-log space to extract the power-law coefficients presented in Eq.~\eqref{eq:decoupling_timescale}. Figure~\ref{fig:power_law_fit} shows the accuracy of this fit against the numerical data.
The corresponding best-fit parameters and their formal $1\sigma$ uncertainties are summarized in Table~\ref{table:fit_params}. These formal uncertainties are much smaller than the theoretical uncertainty associated with the model assumptions and the adopted parameters. 
\begin{table}[ht!]
\caption{Best-fit parameters for the decoupling timescale.}
\label{table:fit_params}
\centering
\begin{tabular}{l r c}
\hline\hline
Parameter & Value & $1\sigma$ Uncertainty \\
\hline
    Normalization & $\expTauNought$ & $\expTauNoughtErr$ \\
    Bulk metallicity exponent & $\expZBulk$ & $\expZBulkErr$ \\
    Opacity exponent & $\expOpacity$ & $\expOpacityErr$ \\
    Mass exponent & $\expMp$ & $\expMpErr$ \\
\hline
\end{tabular}
\tablefoot{
The normalization constant, $\tauDecouple^0$, represents the pre-factor of the fit corresponding to the reference model. We note that the true physical uncertainty is expected to be significantly larger than that of the formal numerical fit.
}
\end{table}
\FloatBarrier
\section{Luminosity evolution}\label{sec:luminosity_evolution}
\begin{figure}[h!]
\centering
\includegraphics[width=1\columnwidth]{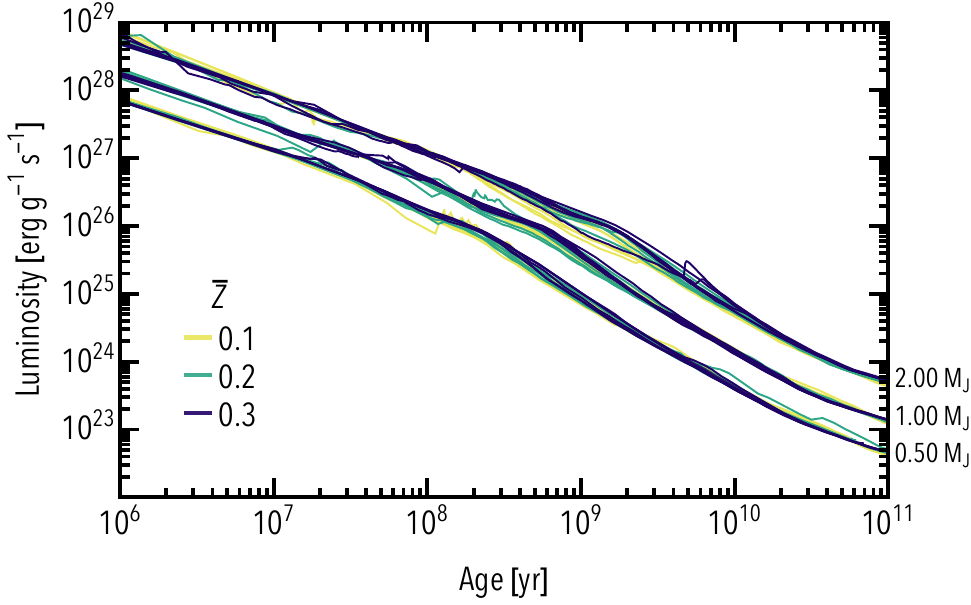}
\caption{Luminosity evolution of the simulations shown in Fig.~\ref{fig:radius_evolution}. The three families of curves correspond to the planetary masses of \qty{0.5}{\MJ}, \qty{1.0}{\MJ}, and \qty{2.0}{\MJ}, indicated on the right.}\label{fig:luminosity_evolution}
\end{figure}

As discussed in Sect.~\ref{sec:observational_implications}, the cooling tracks of gas giants are sensitive to their internal structure. 
The luminosity of gas giants at a given age depends on various physical and chemical properties, including primordial entropies, opacities, composition gradients, and, towards the brown dwarf regime, deuterium burning \citep[e.g.,][]{Burrows_1997,Baraffe_2002,Fortney_2008,Spiegel_2012,Marleau_2014,Muller_2021,Sur_2026}.
Luminosity differences at a given planetary mass and age could be used to constrain the planet formation mechanism, most prominently by differentiating between ``hot starts'' and ``cold starts'' traditionally   associated with the disk instability and core accretion paradigms, respectively \citep[e.g.,][]{Marley_2007, Mordasini_2013, Marleau_2014}. However, this paradigm has been questioned by subsequent studies demonstrating that core accretion can also produce hot starts \citep[e.g.,][]{Berardo_2017, Cumming2018}. In addition, it was shown that the planetary cooling also strongly depends on the assumed core mass in the core accretion framework \citep[e.g.,][]{Mordasini_2013}. 
Given the uncertainties in the parameters mentioned above, relying solely on luminosity to constrain the formation history is extremely challenging. Ultimately, the planetary luminosity is not strictly a function of only the mass and age.
To illustrate this, Fig.~\ref{fig:luminosity_evolution} shows the luminosity evolution of the simulations from Fig.~\ref{fig:radius_evolution}.

We reproduce the aforementioned degeneracy within our numerical experiments, demonstrating that a planet's bulk metallicity (and consequently, its radius) cannot be uniquely inferred from its observed luminosity.
We observe considerable scatter, particularly during the entropy-divergent phase, highlighting the limitations of simple luminosity--mass or luminosity--radius relations. Furthermore, this degeneracy is intrinsic. Even if the planetary mass is measured independently, the internal composition gradient (and the resulting bulk metallicity) can still heavily alter the thermal evolution. Hence, directly imaged planets remain a challenging population to characterize.
\par
The transient spikes observed in the luminosity evolution are a result of rapid convective mixing. Numerically, a region with a composition gradient can become convectively unstable between two timesteps. Given that (1) dredging up material requires energy and (2) the dredged-up material might be high in thermal energy, these relatively sudden mixing events create the spikes observed in the luminosity.
\section{Importance of equilibrium temperature}\label{sec:equilibrium_temperature}
\begin{figure}[ht!]
    \centering
    \includegraphics[width=1\columnwidth]{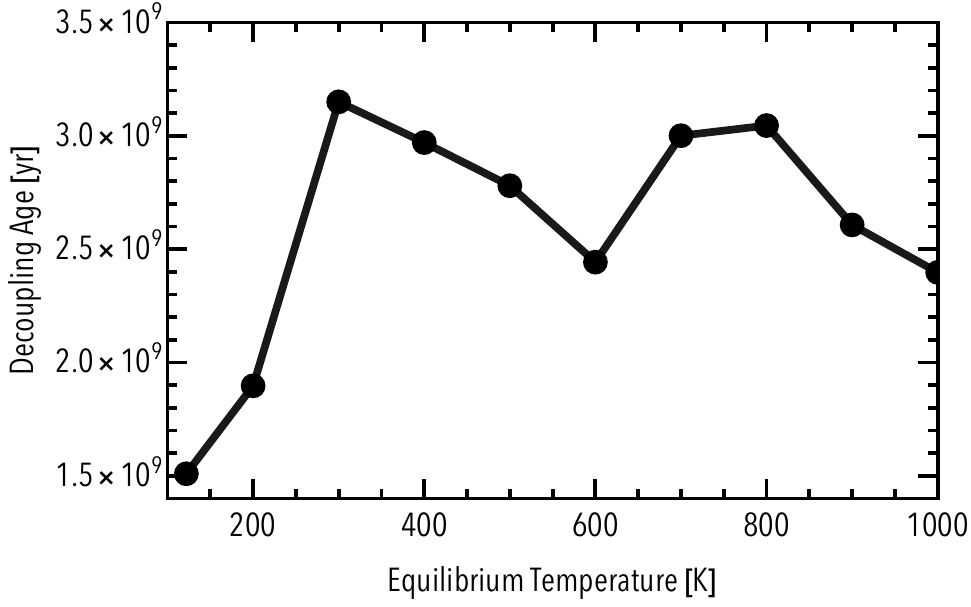}
    \caption{Decoupling age versus equilibrium temperature for a \SI{1}{\MJ} planet with a bulk metallicity of $\Zmean = 0.1$.}\label{fig:equilibrium_temperature}
\end{figure}

Figure~\ref{fig:equilibrium_temperature} shows the computed decoupling age as a function of assumed equilibrium temperature, restricted to $T_\mathrm{eq} \le \SI{1000}{\K}$. 
We impose a limit of $\SI{1000}{\K}$ because our simulations do not incorporate anomalous heating mechanisms (radius inflation). Beyond $\sim\SI{1000}{\K}$, neglecting these processes leads to inaccurate cooling histories and thus planetary radii.
\par
Within this physically robust domain, the derived decoupling ages scatter between about \SI{1.5}{\Gyr} and \SI{3.3}{\Gyr}. The relationship is non-monotonic; the computed age generally increases from a local minimum approaching the zero-Kelvin limit, but exhibits significant scatter across the parameter space.
\par
This complex behavior arises from competing physical effects. On one hand, higher irradiation levels prolong the cooling timescale, which drives the overall upward trend in the decoupling age. On the other hand, higher equilibrium temperatures push the radiative-convective boundary to deeper pressures. Because we evaluate the planetary radius at \SI{1}{\bar}, these varying atmospheric structures introduce an additional offset that interferes with the interior contraction driven by the composition gradient. Under certain conditions, this atmospheric expansion may potentially even mask the radius contribution of the composition gradient.
\par
Ultimately, within the bounds of our framework ($T_\mathrm{eq} \le \SI{1000}{\K}$), the numerically extracted decoupling age changes by a factor of a few. While accurate atmospheric modeling remains necessary to characterize individual exoplanets, the fundamental interior physics for non-inflated gas giants remains: altering the thermal boundary condition simply shifts the timing of the transition.
\section{Helium rain} \label{sec:helium_rain}
\begin{figure}[ht!]
    \centering
    \includegraphics[width=\columnwidth]{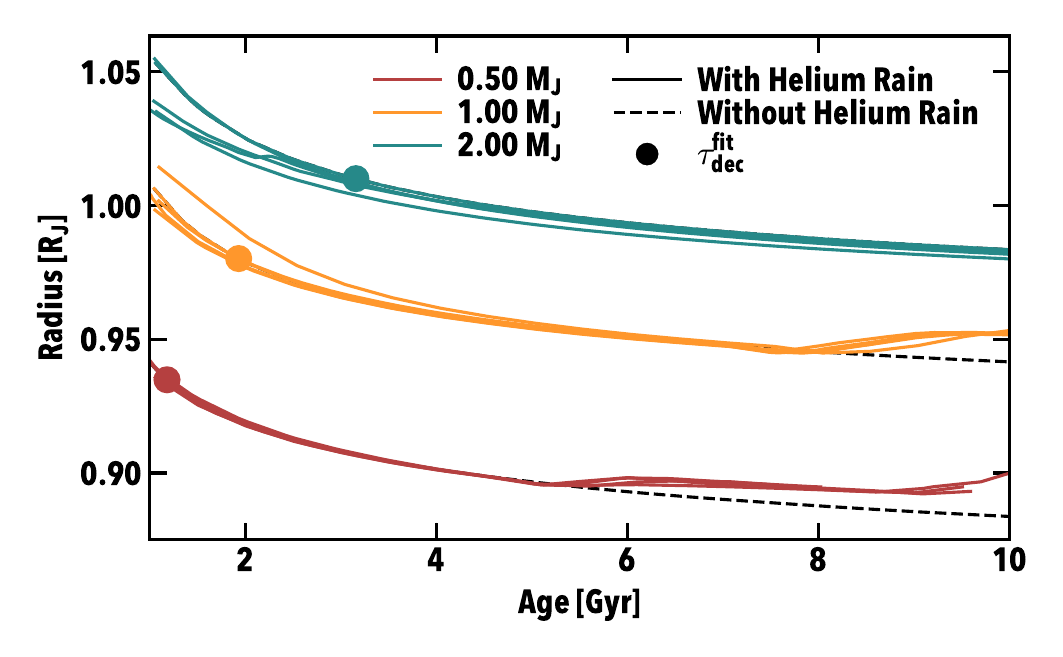}
    \caption{Radius evolution for the subset of simulations from Fig.~\ref{fig:radius_evolution} with $\Zmean = 0.1$ for different planetary masses. For clarity, a single representative curve without helium rain (dashed line) is shown for each mass, as these baseline models follow nearly identical tracks. The filled circles indicate the respective decoupling timescales ($\tau_\mathrm{dec}^\mathrm{fit}$) computed via Eq.~\eqref{eq:decoupling_timescale}.}\label{fig:helium_rain}
\end{figure}

To model the effect of helium rain on the evolution, we used the \texttt{instant\_rain} algorithm introduced by \citet{Knierim2025a}. This algorithm computes the immiscible regions of a \mesa model and instantaneously transfers the immiscible helium fraction into deeper layers \citep[for details, see][]{Helled_2025}.
A determination of these immiscible regions requires an accurate H-He phase diagram. Currently, despite ongoing efforts on both the theoretical and experimental fronts, the phase diagram of H-He demixing remains uncertain. To alleviate some of this uncertainty, evolutionary models typically shift existing phase diagrams by a constant temperature offset to reproduce the present-day atmospheric helium abundance of Jupiter \citep[e.g.,][]{Howard2024}.
In this work, we employed the H-He phase diagram computed by \citet{Schottler_2018} shifted by \SI{325}{\K}, a value that reproduces present-day Jupiter for the best-fitting model of \citet{Knierim2026} and is comparable to shifts introduced by previous studies \citep[e.g.,][and references therein]{Howard2024}.
Figure~\ref{fig:helium_rain} shows the radius evolution of various simulations with and without helium rain for three planetary masses: \SI{0.5}{\MJ}, \SI{1}{\MJ}, and \SI{2}{\MJ}. 

\par
As expected, helium rain increases the planetary radius at late times. This occurs because the precipitating helium droplets release gravitational energy, which increases the total entropy and thereby inflates the planet.
Furthermore, because lower-mass planets have cooler envelopes (and cooler interiors in general due to less compression), helium phase separation sets in earlier for these objects.
Conversely, for the $\Mp = \SI{2}{\MJ}$ simulations, helium rain does not set in within \SI{10}{\Gyr}, making it observationally irrelevant over this timescale.
As discussed in Sect.~\ref{sec:discussion}, the late-stage radius inflation due to helium rain is approximately uniform across the sample of simulations we tested. Because the onset of rainout in these models occurs only after they have already passed the decoupling threshold (Eq.~\eqref{eq:decoupling_timescale}), their radii remain the same for a given mass and bulk composition.
\par
We note, as a theoretical caveat, that the exact timing of this radius increase depends on the assumed H-He phase diagram and its empirical temperature offset. Future constraints on the H-He phase diagram could theoretically shift the onset of helium rain: if an alternative phase diagram were to shift this onset significantly earlier---such that rainout in Saturn-mass planets occurs on a timescale comparable to $\tauDecouple$---it could potentially prolong the time required for the radii to reach the entropy-convergent phase. 
However, we do not expect future phase diagrams to deviate drastically from the physically motivated parameter space adopted here. Therefore, the theoretical framework we present is valid irrespective of helium immiscibility, a topic we hope to address in greater detail in future research.
\end{appendix}
\end{document}